\documentstyle[12pt,aasmanuscript]{article}

\input util.emgdp

\begin{document}

\setlength{\baselineskip}{24pt}

\title{Multiple Outflow Episodes from Protostars: 3-D Models of Intermittent
Jets } \author{Elisabete M. de Gouveia Dal Pino}
\affiliation{University of S\~ao Paulo, Instituto Astron\^omico e Geof\'isico\\
Av. Miguel St\'efano, 4200, S\~ao Paulo, SP 04301-901, Brazil
\\ dalpino@astro1.iagusp.usp.br}\author{ Willy Benz}%
\affiliation{University of Arizona, Steward Observatory and Lunar Plan. Lab.\\
Tucson, AZ 85721, USA} 

\dates

\begin{abstract}
We present fully three-dimensional simulations of supersonic, radiatively
cooling intermittent jets with intermediate and long variability periods
(that is, periods of the order of or longer than, the dynamical time scale of
the jet, $\tau \geq \tau _{dy}$). Variations of intermediate period elucidate
the formation and evolution of chains of internal regularly spaced radiative
shocks, which in this work are identified with the observed
emission knots of protostellar jets. Variations of long period elucidate the
formation of multiple bow shock structures separated by long trails of
diffuse gas, which resemble those observed in systems like HH111 and
HH46/47. The time variability of the outflow is probably associated with
observed irruptive events in the accretion process around the protostars. In
our simulations, the outflow variations are produced by periodically $%
turning $ $on$ the outflow with a highly supersonic velocity and
periodically $turning$ $off$ it to a low velocity regime. When a supersonic
parcel finds the slow material that has been injected earlier, a double
shock structure promptly develops: a $forward $ shock sweeps up the slow
material ahead of it and a $reverse$ shock decelerates the fast material
behind. The very high density contrast between the fast and slow portions of
the flow causes the $reverse$ shock to be much weaker than the $forward$
shock so that line emission by gas between these shocks is essentially
single peaked.

In the case of velocity variations of $intermediate$ period, we find, as in
previous work, that the shock structures form a train of regularly spaced
emitting features which move away from the source with a velocity close to
that of the outflow, have high radial motions, and produce low intensity
spectra, as required by the observations. As they propagate downstream, the
shocks widen and dissipate due to the expulsion of material sideways to the
cocoon by the high pressure gradients of the postshock gas. This fading
explains the most frequent occurence of knots closer to the driving source.
The head structure of these outflows is very affected by interactions with
the internal knots.

In the case of the $long$ period velocity variability, our simulations have
produced a pair of bow-shock-like structures separated by a trail almost
starved of gas extending for many jet radii in agreement with the
observations. The leading bow shock or working surface developed at the head
of the jet has a similar structure to that of steady jets (e.g., Gouveia Dal
Pino \& Benz, 1993): the dense shell formed from the condensation of the
shock-heated gas also fragments into a clumpy structure and becomes
thermally unstable . Parts of the shell spill into the cocoon forming an
extended narrow plug. The narrowing of the jet head causes some beam
acceleration. The bow shock produced from the second injection propagates
downstream on the diffuse tail behind the leading working surface. It has a
double shock structure similar to the one of the leading working surface and
on average propagates faster. The postshock radiative material also cools
and forms a cold shell whose density is much smaller than that of the shell
of the leading working surface. This result is consistent with observations
that suggest that the emission of the internal working surface is of lower
intensity and excitation than that of the leading working surface.

A brief discussion on the possibility of these time-dependent intermittent
jets to drive molecular outflows is also presented.
\end{abstract}


\section{Introduction}

Most protostellar jets show a bow shock-like structure at the edge of the
beam which is identified as the jet working surface (for reviews see, for
example, Mundt 1988; Reipurth 1989a). As the supersonic jet propagates into
the ambient gas it develops a shock pattern at the head. A detailed
three-dimensional numerical analysis of this shock pattern for continuous
flows has been done in a previous paper (Gouveia Dal Pino \& Benz 1993a,
hereafter Paper I). The beam outflow is decelerated in a jet shock or Mach
disk, whereas the impacted ambient material is accelerated by the forward
bow shock. In some jets ({\em e.g.}, HH46/47 and HH 111), there is clear
evidence of two or more such bow shock structures separated by a trail of
diffuse gas for many jet radii ({\em e.g.}, Hartigan, Raymond \& Meaburn
1990, Reipurth 1989b). These multiple bow shocks have been interpreted as
evidence for distinct outflow episodes from the jet source ($e.g.$, Reipurth
1989b, Raga et al. 1990, Reipurth and Heathcote 1991, Gouveia Dal Pino and
Benz 1993b, c).

Usually, along the jet beam, there is a chain of rather regularly spaced
emission knots with large proper motions ($\sim 50-300$ km/s), emitting
mainly low-excitation lines. The knots often become fainter and disappear at
larger distances from the source, as for example, in thr HH34 and HH 111 jet
systems (e.g., Reipurth et al. 1986, Reipurth 1989b, Morse et al. 1992,
Reipurth, Raga \& Heathcote 1992). It is generally believed that the line
emission is produced by shock-heated gas in the jet flow ($e.g.$, Hartigan,
Raymond \& Hartmann 1987).

The precise nature of the mechanism that produces these internal shocks is
still controversial. One of the possibilities is that the shocks are driven
by Kelvin-Helmholtz instabilities excited in the interface between the jet
flow and the surrounding medium (e.g., Birkinshaw 1991). This mechanism has
been examined in detail in two-dimensional (Blondin, Fryxell \& K\"onigl
1990, hereafter BFK) and three-dimensional (Paper I) hydrodynamic
simulations of steady-state radiative cooling jets which showed that, under
the presence of radiative cooling, shocks driven by K-H instabilities are
fainter and less numerous than in adiabatic jets. The radiative cooling
reduces the thermal pressure which is deposited in the cocoon that surrounds
the beam. As a result, the cocoon has less pressure to collimate, drive K-H
instabilities and thus, reflect internal shocks in the beam. Although this
mechanism remains viable, it probably plays a secondary role in the
production of internal shocks.

A more attractive possibility is that the knots (like the multiple bow
shocks) could be the product of time variations in the ejection mechanism
that produces the jet. Strong support for this argument comes from recent
high resolution observations of the HH34 and HH111 jets (Morse et al. 1993)
which, based on the morphology, spectra, and kinematics of the knots, show
evidence of a velocity-variable outflow with multiple ejections. Also, long
slit spectra of HH83 and HH46/47 show that the radial velocity of the
emitting material increases with increasing distance from the source. In the
case of HH46/47, abrupt variations of this velocity have been detected ($e.g$%
., Reipurth 1989a, b). These data are consistent with variable jet ejection
velocity where the knots would be the products of multiple outflow episodes.

The time variability of the outflow is probably associated with irruptive
events in the accretion process around the protostar ($e.g.$, Dopita 1978,
Reipurth 1985). In fact, some energy sources of protostellar jets and HH
objects, as for example the sources of HH57 and HH28/29 ($e.g$., Reipurth
1989b), have been seen to erupt into FU Orionis outbursts. These episodes
could produce massive ejections along the stellar jets every 100-1000 years
(e.g., Hartigan \& Raymond 1992; Hartmann, Kenyon \& Hartigan 1993),
suggesting that the driving sources can change on timescales shorter than
the dynamical timescale of the HH objects.

The possibility that internal knots could emerge from variations in the
ejection mechanism was first proposed by Rees (1978) in the context of
extragalactic jets. He showed that time variations of the supersonic
ejection velocity would produce internal shocks that would travel along the
beam. Later, Wilson et al. (1984) performed two-dimensional numerical
simulations of pulsating adiabatic jets and showed that the internal shocks
evolve to two-shock $internal$ $working$ $surface$ structures. More
recently, Raga et al. (1990) developed an one-dimensional analytical model
employing the Burgers' equation of motion, which neglects the gas pressure
gradient forces in the beam to describe the evolution of high Mach number
flows from time-dependent sources . Their results qualitatively confirmed
the predictions of previous work. Pursuing this idea further, Kofman \& Raga
(1992, hereafter KR), assuming a constant jet density, and Raga \& Kofman
(1992, hereafter, RK), assuming a variable jet density, extended that study
and derived analytic asymptotic solutions of the Burgers' equation for the
jet flow at large distances from the source. They found that the flow would
consist of continuous segments for which the flow velocity would increase
linearly with the distance from the source, separated by discontinuous
velocity jumps which should correspond to the knots or internal working
surfaces propagating down the flow at the mean flow velocity. Using similar
assumptions, Raga (1992) studied the evolution of random perturbations in
the jet velocity. Although their results partially reproduce the
observations, they cannot correctly describe the time evolution of the
density and velocity profiles of the knots as the pressure has been
neglected. In such cases, the motion of the knots is essentially determined
from simple ram pressure balance. This assumption is valid only in the
region close to the jet axis or if the flow is surrounded by a high density
gas at rest. A generalization of their work by Hartigan \& Raymond (1992,
hereafter HR) took into account the pressure effects, performed
one-dimensional numerical hydrodynamical calculations of pulsating cooling
jets, and evaluated the evolution and intensity of the pulses along the jet
axis. To compute the radiative losses of the flow they used a detailed
time-dependent nonequilibrium model.

Although all those calculations qualitatively (and quantitatively) reproduce
some of the observational properties of the protostellar knots ($e.g.$,
proper motions, emission-line intensities, knot spacing), they all have
assumed one-dimensional flows. In general, sudden changes of the jet radius
and the velocity are expected at the position of the internal knots. This
results in a shock structure which is more complex than the one described by
the one-dimensional approach and in which the effects of the thermal
pressure are non-negligible. A multidimensional analysis is then required
for a self-consistent calculation of the structure and dynamics of these
jets from time dependent sources. Very recently, Falle and Raga (1993)
performed two-dimensional numerical calculations of individual internal
working surfaces. In this work, we present fully three-dimensional numerical
simulations of supersonic radiative jets with variable ejection velocity, in
order to examine the formation and evolution of chains of internal knots and
also the formation of the multiple bow-shocks which are observed to be
separated by large portions of very diffuse and almost invisible flow. Fully
3D simulations can provide a complete description of the flow kinematics at
any time in the evolution, subject to the constraints of numerical
resolution.

In order to represent the velocity variability, we assume that the jet is
periodically ''turned on'' with a full supersonic velocity and periodically
''turned off'' to a small velocity. In such ''intermittent'' jets, the
velocity perturbations are of large amplitude and are accompanied by similar
density enhancements (see also HR). To study the development of a chain of
internal knots along the jet, we consider velocity variations with a period
comparable to the dynamical time scale of the jet (which we define in this
work as being $\tau _{dy}\equiv R_j$/$v_j$, where $R_j$ is the jet radius
and $v_j$ is the mean jet velocity). To investigate the formation of
multiple bow shock structures separated by long trails of diffuse gas, on
the other hand, we assume variations with much longer periods than the jet
dynamical time scale.

In Paper I, we presented the first fully three-dimensional simulations of
radiative cooling jets ejected at a constant velocity using the smoothed
particle hydrodynamics technique (SPH). In a subsequent paper (Chernin,
Masson, Gouveia Dal Pino \& Benz 1994; hereafter Paper II), we employed the
same technique to study the dynamics of the transfer of momentum from jets
to ambient molecular clouds in order to examine the correlation between
protostellar jets and molecular outflows. In the present work, we use a
modified version of the code employed in these previous investigations.
Contemporaneously with this work, Stone and Norman (1993a, hereafter SN)
have performed two and three-dimensional simulations of pulsed cooling jets
in which the jet velocity is assumed to undergo sinusoidal small amplitude ($%
\Delta v/v_j<<1$, where $\Delta v$ is the maximum velocity variation) high
frequency oscillations ($\tau <t_j=R_j/c_a$, where $c_a$ is the sound speed
of the ambient medium). In their simulations, the density is pulsed
inversely with the velocity in order to keep the mass flux constant.
Although in our simulations we have assumed higher amplitude, smaller
frequency velocity and density variations ($\Delta v/v_j\simeq 1$, $\tau
\geq t_j>\tau _{dy}$), where both investigations overlap, the results are
qualitatively similar.

The organization of this work is as follows. In \S 2, we discuss briefly the
basic theoretical properties of intermittent jets and describe the
assumptions of our numerical model , including the initial and boundary
conditions, and model parameters adopted for the simulations. In \S 3, we
present the results of our hydrodynamical simulations for jets with both
intermediate and long period velocity variations of high amplitude. Finally,
in \S 4, we summarize our results and discuss their possible applications to
the observed protostellar jets.

\section{Description of the Model}

\subsection{Theoretical Grounds}

The basic properties of jets with continuous injection are described in
Paper I. Jets with supersonic variations in the velocity are expected to
produce shock waves along the outflow as the fast material overtakes the
slower material ($e.g.$, Raga $\ et$ $al.$ 1990, HR). In general, a pair of
shocks must form as the supersonic velocity perturbation steepens. A
downstream (or forward) shock is produced as the perturbation sweeps up the
low velocity material ahead of it and an upstream (or reverse) shock must
decelerate the fast material that collides with the perturbation. This shock
structure resembles the working surface that appears at the head of the jet
where the forward shock is identified with the bow shock and the reverse
shock with the jet shock (e.g., BFK, Paper I).

As we have mentioned earlier, the protostellar jets display a large number
of emitting knots, which in the present work we identify with internal $%
working$ $surfaces$. So, in order to reproduce such a structure, one would
require
\begin{equation}
N_{is}\approx \frac{f\tau _{dy}}\tau \gg 1
\end{equation}
internal working surfaces, where $\tau _{dy}$ is the dynamical time scale of
the flow $\tau _{dy}\equiv R_j/v_j,$ $f=L/R_j,$ $L$ being the total length
of the outflow, and $\tau $ the period of the velocity variability.
Condition (1) could be used, in principle, to predict the variability period
required for producing a given number of internal knots observed in an
outflow.

One-dimensional analysis can give an estimate of the velocity of propagation
of an internal working surface. This velocity can be evaluated by balancing
the momentum flux of the flow upstream with the momentum flux of the flow
downstream of the working surface. If the velocity upstream of the working
surface is $v_u$ and the velocity downstream is $v_d$, then the propagation
velocity of the internal working surface, $v_{is}$, is given by:
\begin{equation}
\rho _u\left( v_u-v_{is}\right) ^2+p_u\simeq \rho _d\left( v_{is}-v_d\right)
^2+p_d
\end{equation}
where $\rho _u$ and $\rho _d$ are the mass densities up and downstream, and $%
p_u$ and $p_d$ are the thermal pressures up and downstream, respectively.
For highly supersonic flows or under the condition of thermal pressure
balance, $p_u-p_d$ can be neglected and the internal working surface
velocity is then roughly given by
\begin{equation}
v_{is}\simeq \frac{\beta v_u+v_d}{1+\beta }
\end{equation}
where $\beta =(\rho _u/\rho _d)^{1/2}$. We see that for $\beta =(\rho
_u/\rho _d)^{1/2}\gg 1$
\begin{equation}
v_{is}\simeq \left( 1-\beta ^{-1}\right) v_u
\end{equation}
We note that eqs.(3 and 4) must be satisfied only in the region close to the
axis of symmetry of the jet, as pressure gradient effects must be non
negligible off the axis (see below).

The primary aim of the present work is to study the dynamical evolution of
jets with high amplitude velocity discontinuities of intermediate and long
period which must steepen to form shocks. A jet that varies in velocity is
likely to vary in density as well. As has been argued by HR, without an
understanding of the mechanism that generates the stellar jets, it is not
clear whether the density varies with the velocity or inversely with the
velocity keeping the mass flux constant (as has been assumed by SN). In the
simulations of ''intermittent'' jets presented here, both the fast flow
portions (ejected in the ''turning on'' phases) and the slow portions
(ejected during the ''quiescent'' phases) are ejected with the same initial
density. However, during the quiescent phase, a trail of very diffuse gas
develops behind the previous fast portion of flow. As a consequence, when a
new fast flow portion emerges from the inlet, it finds a downstream slow
flow portion of low density. A similar scenario has also been investigated
in some of the one-dimensional models of HR and they found that in such
cases, the forward (downstream) shock is much stronger than the reverse
(upstream) shock. (This situation resembles the bullet models of Norman \&
Silk 1979, as we are going to see below.) In the case that the mass-loss
flux is kept constant, the high-velocity flow will be less dense than the
low-velocity one and in this case, the reverse shock is expected to have
larger velocity than the forward shock (see HR, SN).

\subsection{The Numerical Model}

We solve the hydrodynamics conservation equations using the smoothed
particle hydrodynamics technique (SPH). As in Paper I, we use a version of
the three-dimensional Cartesian SPH code (described in detail by Benz 1990,
1991) modified to include the effects of cooling. SPH is a Lagrangean
approach to fluid dynamics which does not require the use of a fixed grid
for the computation of the derivatives but instead employs particles which
track the fluid and move with it. Each SPH particle is characterized by its
position in the 6-D phase-space ($\overrightarrow{r}$,$\overrightarrow{v}$),
its mass, and its specific internal energy. The SPH particles are smoothed
out in space by a spherically symmetric kernel function of width $h$. The
initial values of $h$ were, in general, chosen to be $0.4$ and $0.2R_j$, for
the ambient and the jet particles, respectively.

As in Paper I, the jet and the ambient gas are treated as a single, fully
ionized fluid with a ratio of specific heats $\gamma =5/3$ and an ideal
equation of state $p=u(\gamma -1)\rho $, where $u$ is the internal energy
per unit mass, $p$ is the thermal pressure, and $\rho =n\bar m$ is the mass
density, where $n$ is the particle number density and $\bar m$ is the mean
mass per particle which remains constant in the case of complete ionization.

Radiative cooling (due to collisional excitation and de-excitation, and
recombination) is implicitly calculated using the $local$ time-independent
cooling function evaluated by Katz (1989) for a gas of cosmic abundances
cooling from $T=10^6$ K. The cooling is suppressed below $T\approx 8000$ K
when the assumption of completely ionized flow breaks down and the effects
of transfer of ionizing radiation become important. For $T<8000$ K, the
radiative cooling is mainly due to the formation of molecules and molecular
lines and, therefore, the expected overall cooling will be stronger than
that inferred in our calculations. By not taking into account the effects of
nonequilibrium ionization of the gas or the transfer of ionizing radiation,
we are probably underestimating the cooling rate in some parts of the
postshock regions by as much as an order of magnitude ($e.g.$, Innes,
Giddings, \& Falle 1987). However, a comparison of our results (see also
Paper I) with the calculations of SN (and also with recent multidimensional
calculations for steady flows; Stone \& Norman 1993b), which have included a
nonequilibrium time-dependent cooling, show that the essential dynamical
features do not change under the presence of nonequilibrium ionization
effects.

As pointed out in Papers I and II, in SPH simulations the properties of the
low density regions, as for example, the cocoon that develops between the
swept-up ambient gas and the jet beam, or the region between successive
knots along the beam (see below) are poorly sampled because they do not
contain as many SPH particles as the denser regions. However, one of the
advantages of SPH codes over grid-based codes is not having to calculate
properties of ''empty'' regions. These low density regions may become a
problem only if their pressure is dynamically significant, but we later show
that this is not the case. Another limitation of SPH simulations is its
treatment of turbulence, which is expected to be significant in protostellar
jets since the inferred Reynolds numbers are of the order $Re\simeq $1$%
0^4-10^5$. This difficulty is mainly due to the fact that the numerical
viscosity in the code (see Paper I) may be too dissipative and the particle
size is orders of magnitude larger than the Kolmogorov dissipation scale, $%
\sim $ $d_{cool}$/$Re$ $<$ $R_j/10^4$. However, the main effect of this
viscosity is to suppress numerical instabilities which develop on scales
smaller than the scale fixed by the viscosity law. If the number of SPH
particles is increased, thereby increasing the resolution of the
simulations, smaller scale eddies are seen, but the large scale structures
of the flow remain the same. Thus, our simulations are adequate for studying
the gross properties of the jet system.

The boundary conditions are the same as described in Paper I. All distances
are normalized to the jet radius $R_j$. An initially homogeneous ambient gas
fills the spatial domain which has dimensions of $24$ or $45R_j$ in the
z-axis direction and $-6R_j\leq x,y\leq 6R_j$ in the transverse directions.
A\ jet with a circular cross section is injected at $z=0$ in the middle of
the ''ambient box '' and allowed to propagate. We assume periodic boundary
conditions on the x and y peripheries (see Paper I) and continuitive
boundary conditions at $z=0$ and $z_{max}$. (These conditions are different
from those in the three-dimensional simulations of SN who assumed reflecting
boundaries on the midplanes x=0, y=0 of the jet. In such cases the jet is
symmetric across both planes. This assumption prohibits the development of
nonaxisymmetric modes of dynamical instabilities in the beam.)

The evolution of the system is characterized by the dimensionless
parameters: i) $\eta =\rho _j/\rho _a$ (the ratio between the input jet and
ambient densities); (ii) $M_a=v_j/c_a$ (the initial ambient Mach number,
where the ambient sound speed is given by $c_a=(\gamma kT_a/\bar m)^{1/2}$,
with $T_a$ being the initial ambient temperature and $\bar m\simeq 0.5m_H$
the mean mass per particle for a fully ionized gas of cosmic abundances);
(iii) $k_p=p_j/p_a$ (the input pressure ratio which has been assumed to be
initially equal to unity in all simulations); and (iv) $q_{bs}$ (the ratio
of the cooling length in the postshock gas behind the bow shock (at the jet
head) to the jet radius (see eq. 7 of Paper I), which, for shock velocities $%
v_s>90\ $km/s is related to the ratio measured in the postshock gas behind
the jet shock $q_{js}$ through $q_{js}\simeq q_{bs}\eta ^{-3}$ (see eq. 8 of
Paper I).

The parameters of the simulations were chosen to resemble the conditions
found in protostellar jets. The densities of protostellar jets are not well
determined. Early estimates indicated jet number densities $n_j\equiv \eta
n_a\sim 20-100$ cm$^{-3}$ ($e.g.$, Mundt, Br\"ugel \& Buhke 1987, hereafter
MBB), but recent observations seem to indicate values larger than $10^3$ cm$%
^{-3}$ ($e.g.$, Morse et al. 1992., 1993). The density ratio $\eta $ is even
more uncertain. Observations suggest values $\eta \simeq 1-20$ (e.g., MBB,
Morse et al. 1992, Raga \& Noriega-Crespo 1993). Typical jet velocities have
been determined from proper motions and radial velocities of the knots and
range from $v_j$ $\simeq 100-500$ km/s (e.g., Reipurth 1989b, Reipurth et al
1992), corresponding to jet temperatures $T_j=\eta ^{-1}T_a\simeq 10^4$ K$.$
Typical jet radii are $0.5-4\times 10^{16}$ cm ($e.g$., MBB). In this work,
we adopt for the parameters above the values given in Table 1.

In order to model the time variability, we assumed large velocity variations
such that the maximum velocity change is $\Delta v\sim v_j$ with
intermediate and long periods $(\tau \geq R_j/c_a>\tau _{dy}=R_j/v_j$$)$.
The jet injection velocity at the inlet is varied according to a
''periodic'' step function, so that the jet is periodically ''turned on''
with a supersonic velocity $v_{on}$ for a period $\tau _{on}$, and
periodically ''turned off'' to a small velocity $v_{off}$ for a period $\tau
_{off}$. The velocity change is then, $\Delta v=v_{on}-v_{off}\simeq v_{on}$
which we identify with $v_j$, that is, $v_{on}\equiv v_j$. This is
consistent with observations which indicate that the amplitude of the radial
velocity variations in the emission knots can be very large (e.g., in the
HH46/47 jet) and the knot shock speed can be greater than $100$ km/s
(Reipurth 1989a). The multiple bow shock structures that move into the wakes
of previous ejections, as in HH111, are also expected to be produced by
large amplitude velocity variations. If both the internal knots and the
multiple bow shocks are a product of episodic accretion events, as we are
assuming here, then massive ejections along the stellar jets could be
produced every 100-1000 years (e.g., HR; Hartmann, Kenyon \& Hartigan 1993).
So, for the period of variability, we take values which range from few to
many times the dynamical time scale of the jet ($\tau \simeq (6-40)$$\tau
_{dy}$), corresponding to few hundred years. These values are representative
and are not intended to fit any particular system. Other choices are
possible, for example, SN chose to model systems with smaller amplitude and
higher frequency velocity variations.

\section{The Simulations}

The values of the parameters of our simulations are listed in Table 1. The
first model (I) describes the evolution of an $\eta =10$ cooling jet subject
to high amplitude velocity perturbations of intermediate period. The second
model (II) describes the evolution of an $\eta =3$ cooling jet subject to
high amplitude velocity perturbations of long period. Subsections \S 3.1 and
\S 3.2, present the results of the first and second calculations,
respectively.

\subsection{The evolution of a jet with intermediate period velocity
variations}

Figure 1 depicts the time evolution of the central density contours of an
intermittent jet periodically ''turned on'' with a supersonic velocity $%
v_j=150$ km/s and periodically ''turned off '' with a subsonic velocity $15$
km/s. The ''turning on'' and quiescent periods in this case are both given
by $\tau _{on}=\tau _{off}=(R_j/c_a)\simeq 127$ yrs, corresponding to a
total period $\tau \simeq 254$ yrs. Compared to the jet dynamical time scale
$\tau _{dy}=R_j/v_j,$ $\tau _{on}=\tau _{off}=3\tau _{dy}$. The initial
input parameters are (see model I, Table 1) $\eta =10,$ $n_a=1000$ cm$^{-3}$%
, $R_j=2\times 10^{16}$ cm, $M_j=9.5$ (corresponding to $M_a=3$), and $%
T_j=T_a/\eta =9090$ K. The entire evolution displayed in Figure 1
corresponds to t$\simeq 1387$ yr (or 11$R_j/c_a$). The initial value of the
cooling distance parameter for the leading bow shock at the jet head is
given by $q_{bs}\simeq 7\times 10^{-3}$, corresponding to a bow shock
velocity $v_{bs}\simeq 114$ km/s (see eqs. 2 and 7 of Paper I). The jet
shock (or Mach disk), for which $v_{js}\simeq v_j-v_{bs}\simeq 36$ km/s, has
a cooling length parameter $q_{js}\simeq 0.31(\frac{n_j}{1000}cm^{-3})^{-1}(
\frac{v_j}{45}km/s)^{-4.7}(\frac{R_j}{2\times 10^{16}}cm)^{-1}\simeq 9\times
10^{-2}$ (evaluated from eq. 2.13 of Falle \& Raga 1993, which is valid for
shock velocities $v_s<90$ km/s). These values imply that both shocks in the
jet head cool very fast ($q_{bs}$, $q_{js}\ll 1$) and are, therefore, nearly
isothermal. The cold shell formed in the jet head is thus very thin (as
shown in Figure 1) and the head itself resembles a bullet of dense gas
moving through the ambient medium (see also Figure 6 of Paper I).

The parcels of supersonic flow which are injected every $254$ yrs (and keep
flowing for half a period of $127$ yrs), are clearly evident. In each time
interval of Figure 1, a new parcel emerges from the jet inlet (except in
Figures 1E and 1F which show the same emerging parcel). They cause the
formation of pulses which quickly evolve to a chain of shock structures.
Figure 2A gives a closer view of the last snapshot of Figure 1, that is, at $%
t\simeq 1387$ yr. Figure 2B shows the corresponding velocity distribution
map at that time. As the shocks propagate downstream, they tend to widen and
weaken. The number of internal shock structures formed along the outflow ($%
z_{max}=24R_j$) is in agreement with the number expected from eq. 1 ($%
N_{is}\simeq 4$). During the subsonic phases, the jet beam is starved of
material, producing ''gaps'' between the supersonic parcels. We note also
that constant interaction with the evolving internal peaks causes the head
of the jet to have a structure which differs significantly from that of a
steady jet (see Figure 6 of Paper I).

Each internal supersonic parcel develops a pair of shocks: a forward
(downstream) shock which sweeps up the low velocity, diffuse material ahead
of the perturbation and propagates downstream in the jet with a velocity v$%
_{is}$; and a reverse (upstream) shock which decelerates the high velocity
material in the perturbation and propagates with a velocity v$_{rs}\simeq
v_j-v_{is}.$ In order to illustrate in more detail how the discontinuities
of the jet in Figure 1 evolve, Figure 3 shows the axial density along the
symmetry axis (x=y=0) for the six snapshots shown in Figure 1. (The axial
pressure has a similar evolutionary behaviour.) The density evolution of the
leading working surface at the head of the jet is also depicted. We find
that for each internal shock structure a density peak develops between the
two shocks . Due to the long period of the velocity variations, the peaks do
not interact and remain separated by the rarefied material injected in the
subsonic phase. As the peaks travel downstream, they widen and fade and
eventually disappear close to the leading working surface. These effects are
due to the increase in the pressure of the postshock gas which causes both
the separation between each pair of shocks and the expulsion of jet material
laterally to the cocoon and to the rarefied portions of the beam itself.

The lateral widening and fading of the internal shocks is highlighted in
Figure 4 which illustrates the density (solid lines) and pressure (dashed
lines) across the flow as a function of the distance from the source at $%
t\simeq 1387$ yr (or $11R_j/c_a$) for the four internal shock structures of
Figures 1F and 3F. The high pressure and density of the leading working
surface have been clipped in the figure in order to highlight the lower
level features of the internal outflow. As the internal shock structures
propagate downstream, the pressure in the postshock medium pushes the
material to the cocoon that surrounds the beam. As a consequence, the
density and pressure of the shocked material decrease. This result is in
agreement with the results of SN. We note that the density and pressure are
very low in the cocoon. Thus the undersampling of the cocoon by the SPH
particles is acceptable, since the cocoon has little dynamical effect. The
material which the internal shocks push into the cocoon produces wakes that
can be seen in Figure 1, and gives some of these structures the appearance
of internal bow shocks. This effect has been predicted by Raga et al. (1990)
and is also clearly evident in the two-dimensional maps of SN.

Figure 5 shows the position as a function of time of the leading working
surface and the internal shock structures along the symmetry axis of the
jet. The shocks labeled with numbers 6 and 7 have emerged from the jet inlet
after $t=12R_j/c_a=1524$ yr when the leading working surface had already
left the computation domain.

The internal shocks move downstream with an average velocity $%
v_{is}/c_a\simeq 2.7,$ or in other words, with nearly the jet velocity in
the supersonic phase $v_u\simeq v_j=3c_a=150$ km/s. This result is in
agreement with eqs. (3) and (4) for a density ratio of the fast high density
material (upstream of the supersonic discontinuity) to the low density
material (downstream the discontinuity) $\rho _u/\rho _d\gg 1$, as is the
case (see $\S $ 2.1). We note that in the present study, due to the high
density ratio between the fast and the slow portions of the flow, the
propagation velocity of the forward shock, which is of the order of $v_j$,
is much larger than that of the reverse shock $v_{rs}\simeq 0$ (see eq. 3).
So, the forward shock is much stronger than the reverse shock. Similar
behaviour occurs in the bullet models (Norman \& Silk 1979), and has been
detected in one of the one-dimensional models investigated by HR in which
the velocity perturbations are accompanied by similar density enhancements.
The domination of the strong forward shock implies that line emission occurs
in a single peak behind this shock.

Figure 5 also indicates that the average propagation velocity of the leading
bow shock at the head of the jet is smaller than the velocity of the
internal shocks and is initially in rough agreement with the value estimated
from eq. (2) of Paper I ($v_{ws}/c_a\simeq 2.2$). The variations in the
propagation velocity of the leading working surface indicated by the figure
between $t/127 yr \simeq 7 - 9 $ (which result a minimum propagation
velocity $v_{ws}/c_a\simeq 0.9$ at $t/127 \simeq 8.3$) are caused by a
global thermal instability in the leading bow shock. This effect has been
studied in detail in Paper I. High velocity radiative shocks, when subject
to small variations in their shock velocity, oscillate between radiative and
nearly adiabatic shock phases. Accompaning these variations in the shock
velocity, are oscillations in the density and pressure of the shocked
material. These oscillations are clearly evident in Figure 6 which depicts
the time evolution of the axial density (Figure 6A) and pressure (Figure 6B)
of the shocked cooled material at the head of the jet (labeled $ws$). The
maximum of the density and pressure oscillations corresponds to the
radiative jet shock phase while the minimum of the oscillations correspond
to the nearly adiabatic jet shock phase (see Paper I).

Figure 6 also shows the density and pressure evolution of the internal knots
along the axis of the jet. Due to their short lifetime compared to the
leading working surface, we do not detect any evidence of density and
pressure oscillations in these internal shocks, Figure 6 shows that all
internal shocks fade after $\sim (5-6)R_j/c_a$, which is of the order of one
half the period of the oscillations of the leading working surface. The
strong cooling in the postshock gas keeps the internal shocks effectively
isothermal. As a consequence, they produce large compressions and low
postshock temperatures ($T\simeq 10^4$K) before they vanish. Figure 6A
indicates that the maximum density contrast across the internal shocks is n$%
_{is}$/n$_a\simeq 30$, while the maximum density contrast at the leading
working surface is $n_{ws}$/$n_a\simeq 65.$ This implies that the emission
from internal shocks is of smaller intensity and excitation than the
emission from the head, in agreement with the observations. Adiabatic
pulsating jets show smaller compression ratios in the internal shocks
(Wilson 1984).

\subsection{The evolution of an intermittent jet with long period velocity
variations}

Figure 7 depicts the central density evolution and the corresponding
velocity distribution of an overdense cooling jet with a long-period
variability. The input parameters of this simulation (model II) are given in
Table 1. This simulation has been particularly motivated by the observations
of systems like HH111, which shows evidence of two bow shocks separated by
many jet radii on each side of the source, moving away from the source, and
leaving only a faint and diffuse trail of gas. These observations suggest
that if the multiple bow shocks are due to multiple outflow episodes from
the source then they could correspond to long-period velocity variations
like the one presented in Figure 7.

The jet in Figure 7 was initially ''turned on'' with a supersonic velocity $%
v_j=500$ km/s (which corresponds to a jet Mach number $M_j=30$ and an
ambient Mach number $M_a=17.3$) for a period $\tau _{on}$which was much
greater than the dynamical time scale of the jet ($\tau _{dy}=R_j/v_j$): $%
\tau _{on}=1.3(R_j/c_a)=22.5\tau _{dy}=143$ yr (for $R_j=10^{16}$ cm). It
was then turned off for a period $\tau _{off}=1(R_j/c_a)=110$ yr to a
''quiescent'' phase with a weakly supersonic velocity of $29$ km/s ($M_j=1.7$%
). After that period, the jet was returned to the highly supersonic phase in
which it continued until the end of the calculation $t\simeq
3.4(R_j/c_a)\simeq 587(R_j/v_j)\simeq 374$ yr. Only one internal working
surface developed over the whole space domain of the computation ( $%
z_{max}\simeq 45$ $R_j)$. This result is consistent with eq. 1 which
predicts a number of internal working surfaces $N_{is}\simeq 1$ for this
case. Figure 7A depicts the evolution of the first highly supersonic parcel
that gave origin to the leading working surface. Figure 7B shows the second
fast parcel emerging on the trail of diffuse gas formed behind the first
parcel during the slow, quiescent phase. In Figure 7C, the second supersonic
parcel has developed an internal working surface with a bow shock structure.
The evolution of the jet before the time sequence illustrated in the figure
is very similar to the evolution of the jet of Figure 4 of Paper I which has
similar input parameters. For this reason, we do not present here its early
evolution.

The input parameters (see Table 1) correspond to an initial cooling length
parameter for the leading bow shock at the head of the jet $q_{bs}\simeq 15$
and an initial bow shock velocity $v_{bs}\simeq 317$ km/s (see eqs. 2 and 7
of paper I). The leading jet shock has an initial cooling length parameter $%
q_{js}\simeq q_{bs}\eta ^{-3}\simeq 0.56$ (see eq. 8 of Paper I) and a jet
shock velocity $v_{js}\simeq v_j-v_{bs}\simeq 183$ Km/s. These values imply
that the shocked ambient material at the jet head is essentially adiabatic ($%
q_{bs}\gg 1$) while the shocked jet material is radiative ($q_{js}<1)$.
Thus, the gas that has accumulated at the head of the jet forming a dense
shell of cold material consists of the shock-heated $jet$ material that has
cooled by radiation. We note from Figure 7 that the shell at the head of the
jet is clumpy and irregular. This structure is very similar to those found
in three-dimensional simulations of steady jets (see, for example, Figure 4
of Paper I), and occurs because the dense shell becomes dynamically unstable
to the Rayleigh-Taylor instability. Part of these pieces into which the
shell breaks spill out to the cocoon forming, together with the shell, an
elongated narrow plug of cold gas in the head as shown in Figure 7 (see
Paper I for more details).

The velocity of propagation of the leading bow shock increased with time.
This acceleration is directly related to the narrowing of the jet beam at
the head and is common with steady jets (e.g., Paper I). According to eq. 2
of Paper I, the velocity of the leading bow shock is given by $v_{bs}\simeq
v_j[1+(\eta \alpha )^{-1/2}]^{-1}$, where $\alpha =(R_j/R_h)^2$ is the ratio
between the radius of the jet beam and the radius of the jet head. In Figure
7, the velocity of the leading bow shock is initially given by $v_{bs}\simeq
317$ km/s and is consistent with the equation above with $\alpha =1.$ At the
end of the integration, in Figure 7C, $\alpha $ has increased to $\alpha
\simeq 4$ and $v_{bs}\simeq 412$ km/s in fair agreement with the prediction.

A cavity is formed behind the first supersonic parcel during the
''quiescent'' phase (Figure 7B). Later, it is partially filled by jet
material injected in the second highly supersonic phase (which also pushes
the low velocity gas injected in the earlier quiescent phase) and by
material of the cocoon, which is composed of the remnants of the dense shell
of the head. The three little blobs seen between $z=20$ and $30R_j$ in
Figure 7C, are local density concentrations that were part of these remnants.

During the quiescent phase, jet material is injected in the inlet at a low
but weakly supersonic velocity ($M_j=$1.7). It thus develops a very weak
shock as it propagates dowsntream on the diffuse trail left by the first
highly supersonic parcel. This shock front is then overtaken by the second
highly supersonic parcel which starts to be ejected at $t=2.3R_j/c_a\simeq
253$ yr. A double shock structure (see Figure 7C), similar to the one at the
leading working surface, quickly develops. A forward bow shaped shock is
formed as the parcel sweeps up the low velocity material ahead of it and a
reverse weaker shock is formed when the supersonic material coming behind is
decelerated at the discontinuity. (Eventually, a triple shock structure can
be seen on the plots of the axial pressure. The triple shock is a
consequence of the superposition of the weak shock associated with the slow
phase and the double shock that originated with the injection of the second
highly supersonic parcel.) The propagation velocity of this internal working
surface (which corresponds to the internal bow shock speed) is initially
given by $v_{is}\simeq 383$ km/s which corresponds approximately to the
value predicted by eq. 3 for a density ratio $\beta ^2\simeq \eta =3,$ which
is appropriate since the slow material has adjusted to a density close to
that of the ambient gas). This internal working surface experiences some
acceleration as it propagates downstream through the cavity behind the
leading parcel. By the time depicted in Figure 7C, the propagation velocity
has increased to a value close to the jet velocity $v_{is}\simeq 500$ km/s
and the corresponding reverse shock velocity has decreased to $v_{rs}\simeq
v_j-v_{is}\simeq 0$. Later, the velocity of the internal working surface
decreases again as it finds increasing resistance of the material of the
tail behind the leading working surface.

As in Figure 1 (see also Figures 4 and 5 of Paper I), the density of the
shell of the leading working surface varies with time as the radiative shock
becomes thermally unstable (see Paper I). The density in the $z-$axis varies
from $\simeq 1070/n_a$ (in Figure 7A), to $\simeq 1110/n_a$ (in Figure 7B)
and fades to $\simeq 950/n_a$ (in Figure 7C). The corresponding temperature
of the cold shell is $T\simeq 10^4$ K.

The internal working surface which on the average propagates faster than the
leading working surface has a weaker shock structure. The reverse (upstream)
shock is more radiative than the forward (dowstream) shock. This
shock-heated material then cools and forms a cold shell whose maximum
density on the axis remains approximately constant over the time interval
depicted in Figures 7B and C, with an average value $\simeq 6.8n_a,$
corresponding to a temperature $T\simeq 1.4\times 10^4$ K. In Figure 7B, the
axial density of the postshock cold material is 426 cm$^{-3}$, and in Figure
7C, it is 435 cm$^{-3}$. This result is consistent with the expectation that
the emission of the internal working surface will be of lower intensity and
excitation than that of the leading working surface, as required by most
observations. Later, when the leading working surface has left the domain of
integration, the density of the material of the secondary working surface
increases to $\simeq 10n_a.$ Unlike the case of the internal knots formed
with small period variability (Figure 1), no fading of the internal working
surface has been detected with long period variability. Figure 7C also shows
that the beam associated with the second working surface experiences some
collimation caused by the high pressure, shock-heated ambient gas that had
been deposited in the cocoon surrounding the beam. This collimation pinches
the beam and drives a weak internal shock at $z\simeq 3R_j$.

It is interesting to note that the separation between the two working
surfaces in Figure 7C, $\sim 29.5$ $R_j=2.95\times 10^{17}$ cm, is
consistent, for example, with the separation between the two pairs of bow
shocks observed on both sides of the HH111 jet: the western pair (bow shocks
P and V) has a separation of $\sim 5\times 10^{17}$ cm and the eastern pair
(bow shocks X and Y) has a separation $\sim 1.8\times 10^{17}$ cm; Reipurth
(1989b). Our model is also consistent with the observed higher intensity and
excitation of the leading bow shock structure of the western jet relative to
the internal bow shock. Of course, the input parameters do not match this
particular jet system exactly, our results agree only qualitatively with the
observations.

\section{Discussion and Conclusions}

In this work, we have presented fully 3-D hydrodynamical simulations of
overdense, radiative cooling, intermittent jets with high amplitude velocity
variations ($\Delta v\simeq v_j)$ of intermediate ($\tau \simeq \tau
_{dy}=R_j/v_j)$ and long ($\tau \gg \tau _{dy})$ variability periods. These
simulations are qualitatively in agreement with the observed emission knots
of protostellar jets ($\tau \simeq \tau _{dy})$ or with the observed
multiple bow shock structures separated by long trails of diffuse gas ($\tau
\gg \tau _{dy})$. However, our simplified treatment of the radiative cooling
of the gas prevents us from performing a detailed comparison of our models
with the radiation from observed protostellar jets. In particular, the
density contour maps presented here while describing reasonably well the
expected gas distribution do not necessarily correspond to the observed
emission line images. For example, the elongated dense plugs that can
eventually develop at the head of the jet (as in model II; see below) as a
consequence of the disruption and spilling to the cocoon of the cold shell
might not be seen in the images if the plug has become mostly neutral and
therefore unable to strongly radiate. None the less, with the help of
the density contour
maps, we can delineate the basic structural characteristics and dynamics of
the protostellar jets.

We found that the high amplitude supersonic velocity variations of
intermediate period (model I) quickly evolve to form a chain of regularly
spaced radiative shocks which have large proper motions, high radial
velocities, and low intensity spectra. All these characteristics are in
agreement with the observed properties of the knots of protostellar jets and
have also been noted independently on the models of HR and the SN.

The number of internal shock structures that develop along the jet is
roughly given by eq. 1 which relates the dynamical time scale of the jet to
the period of velocity variability. When a supersonic parcel impacts the
slow rarefied material that has been injected in an earlier quiescent phase,
a $forward$ shock quickly forms (as it sweeps up the slow material) and
propagates downstream at $v_{is}\simeq v_j$, the jet velocity in the highly
supersonic phase. The velocity variations in our simulations are accompanied
by high density contrasts between the fast and the slow portions of the
flow. This causes the $reverse$ shock (where the discontinuity decelerates
the fast material coming from behind) to be much weaker than the $forward$
shock (see \S 3.1). As a consequence, the emission behind the discontinuity
must be essentially single peaked, instead of double peaked as in the SN and
some of the HR models that have assumed mass flux conservation. In those
cases, the reverse shock is stronger and so is the emission behind it. Our
internal shocks resemble the bullets modeled by Norman \& Silk (1979).

The efficient radiative cooling of these internal shock structures causes
them to become dense and cold. As they propagate away from the source, they
widen and eventually fade close to the jet head. This is caused by the high
pressure gradient developed in the postshock gas which pushes the material
sideways to the cocoon and increases the separation between the pair of
shocks that bound each density peak . Since the internal shock structures
propagate at nearly the supersonic jet speed they are weaker than the shock
structure at the head of the jet. The maximum density behind the internal
shocks is smaller than the maximum density of the cold shell at the head of
the jet. This indicates that these internal shocks are of lower intensity
and have lower excitation emission spectrum than the jet head, as required
by the observations. Also, the widening and fading of the internal shocks as
they propagate downstream implies that knots must be more common close to
the driving source. This result is also in agreement with the observations.

The larger amplitude of the velocity variations in our simulations compared
to those of SN caused an steepening of the discontinuities into stronger
shocks with higher postshock pressures much closer to the jet inlet than in
their models. As a consequence, there is a more efficient expulsion of the
shocked material sideways to the cocoon which, in turn, causes a faster
widening and fading of the knots. Also, due to the longer period of the
velocity variations compared to the SN models, the internal shock structures
in our simulations remain independent of one another.

In the case of the long period velocity variability (model II), our
simulations have produced a pair of bow shaped working surfaces separated by
a long trail of very diffuse gas found in systems like the HH111 jet. A
leading working surface develops at the head of the jet when the first
highly supersonic portion of flow is injected into the ambient medium.
Contrary to the jet of model I (whose head structure is modified by
interactions with the internal knots), the head structure of the jet of
model II is very similar to that of steady jets (Paper I). A dense shell is
formed by the radiative cooling of the shock-heated jet gas. As in steady
jets, the shell disrupts into pieces which resemble the knotty structure
observed in many HH objects (e.g., HH1, HH2, HH19, HH12) (e.g., Bohm \& Solf
1985; Strom, Strom \& Stocke 1983; Mundt et al. 1984). The shell also
suffers density variations with time which are due to the development of
global thermal instabilities of the radiative shock. As we have argued in
Paper I, such density variations may have important effects on the emission
pattern of the HH objects and may explain, in particular, the brightness
variability that is sometimes observed. After the disruption, parts of the
clumpy shell spill out to the cocoon forming, with the shell, an extended
plug. The narrowing of the jet beam at the head causes some acceleration of
the jet system.

After the quiescent phase (which follows the injection of the first
supersonic portion of flow) in which the jet material is injected at a low
velocity, a second portion of highly supersonic flow is introduced in the
system and develops a second working surface which propagates down the flow.
A trail of very diffuse gas separates both working surfaces by many jet
radii, in agreement with the observations of, for example, the HH111 and
HH47 jets. This second working surface has a double shock structure similar
to that in the leading working surface: a forward bow shaped shock is formed
as the low velocity material is swept up and a weaker reverse shock is
formed when the supersonic material coming from behind is decelerated at the
discontinuity. The initial velocity of propagation of this working surface
is smaller than the jet injection speed, but the beam accelerates as it
propagates downstream through the cavity behind the leading working surface,
and the velocity increases close to the jet injection speed. Later, as the
working surface encounters the increasing resistance of the material of the
wakes behind the leading working surface, its velocity decreases again.

The postshock radiative material also cools and forms a cold shell in this
second working surface whose density is much smaller than that of the shell
of the leading working surface. This result is consistent with the
observation that the emission of the internal working surface is of lower
intensity and excitation than that of the leading working surface.

Of course, since we did not choose input parameters that match exactly
particular jet systems, our results agree only qualitatively with the
observations. By adjusting the amplitude and period of the velocity
variations, models might be generated which fit the morphology of the knots
and multiple bow shock structures of protostellar jets.

It is well known that many young stellar objects are associated not only to
the highly collimated, fast optical jets investigated in this work, but also
to less-collimated, slower molecular outflows detected at radio wavelenghts
(e.g., Rodr\'iguez 1989). Several authors have recently explored models in
which the molecular outflows are driven solely by the collimated jets (e.g.,
Lizano et al 1988, Stahler 1993, Masson \& Chernin 1993, Raga \& Cabrit
1993, Raga et al. 1993, Paper II). In Paper II, where we have investigated
the momentum transfer process between steady-state jets and the environment,
the 3-D simulations indicate that the protostellar jets primarely transfer
momentum to the ambient medium at the working surface. In particular, we
have found that the observed extremely high velocity (EHV) CO features of
the molecular outflows (which may also appear associated to H$_2$ emisson
knots, e.g., Bachiller et al. 1990) can be formed in the swetp-up post-shock
gas at the bow shock. On the other hand, Raga \& Cabrit (1993) have proposed
that the molecular emission could be identified with the swept-up
environmental gas that fills the cavity formed behind the internal working
surfaces of a time-dependent ejected jet. In the case of the jet with
intermediate variability period studied in the present work (model I), there
is no evidence that the ambient material pushed aside by the internal bow
shocks have had time to fill the cavities behind them (see Figures 1 and 2).
The swept-up ambient material cools at the edge of the cavities. But, our
jet with long variability period (model II, Figure 7) clearly shows that the
ambient gas swept up by the leading working surface eventually refills the
cavity formed behind it, in agreement with Raga \& Cabrit's model. Also,
Raga et al. (1993) explored a model in which the molecular outflows
correspond to the turbulent envelopes of mixed jet and ambient material
developed around time-dependent jets. This scenario is believed to be more
appropriate for the case of very evolved outflow systems. According to the
observations, the molecular gas can typically extend over transverse sizes
which are about 20 times wider than the jet diameter. Our results for the
evolved system of model II (Figure 7) indicate that the envelope
does not seem to much exceed $\approx $6 jet diameters.
However, this ''narrowing''
tendency of the distribuition of the surrounding gas in our simulations can
be partialy due to the assumed periodic boundaries
on y and x peripheries
of the ambient box.  Weak reflecting waves may develop on these boundaries
on the late stages of evolution of a system causing some artificial
collimation  of the ambient envelope
(see discussion in Paper I).

As in our previous investigations (Paper I, Paper II), in this study we have
assumed a history-independent optically thin radiative cooling function to
comput the losses of a fully ionized flow. By not following the
history-dependent effects of nonequilibrium ionization of the gas or the
transfer of ionizing radiation, we possibly underestimated the cooling rate
in some parts of the postshock regions by as much as an order of magnitude
(e.g., Innes, Giddings \& Falle 1987). However, the inclusion of a detailed
cooling evaluation, like the one performed in the one-dimensional
calculations of HR, would require a substantial increase in computer power.
Although future work should take into account those effects, we expect that
the gross dynamical features obtained in the present analysis will not
change. This expectation is supported by the comparison of our results (see
also Paper I) with the calculations of SN (see also their recent
calculations for steady flows; Stone \& Norman 1993b), which included a
nonequilibrium time-dependent cooling.

Finally, other possible origins of the internal shocks in protostellar jets
could be the entrainment of ambient material surrounding the jet or the
interaction of the beam with an inhomogeneous preshock medium. The first of
these processes seems to play a secondary role in protostellar jets. In the
numerical investigation of the momentum transfer processes between
steady-state cooling jets and the ambient medium (Paper II), we found that
the entrainment of ambient gas along the jet beam is important only in low
Mach number $(M_j\leq 3),$ low density ratio ($\eta \leq 3)$ flows which, in
general, is not the case for the typical protostellar jets for which $%
10<M_j<40$ and $\eta $ $>1$. The second possibility is examined in a
forthcoming paper (Gouveia Dal Pino, Birkinshaw, and Benz 1994).

\vspace{1in}

E.M.G.D.P. would like to acknowledge fruitful discussions with M.
Birkinshaw, L. Chernin, and A.C. Raga. We are also very grateful to the
referee for his relevant comments and suggestions. The simulations were
performed on the Workstation HP apollo 9000/720 of the Plasma Astrophysics
Group of the Department of Astronomy of IAG/USP, whose purchase was made
possible by the Brazilian agency FAPESP.

\newpage

\begin{center}
{\bf TABLE 1} \\ Values of the parameters used in the models. \\
\begin{tabular}{||c|c|c|c|c|c|c|c|c|c||} \hline
Model   &  $n_{a}$ (cm$^{-3}$) & $T_{a}$ (K) & $R_{j}$ (cm) & M_{j}& $\eta$ &
$q_{bs}^{*}$ & $q_{js}$ & $\tau_{on}/\tau_{dy}$ & $\tau_{off}/\tau_{dy}$ \\
\hline
I  &  1000  &  $9 \times 10^{4}$  &  $2 \times 10^{16}$  & 9.5 & 10.0 & $7
\times 10^{-3}$ & $9 \times 10^{-2} ^{\dag}$ & 3.0 & 3.0 \\
II  &  60  &  $3 \times 10^{4}$  &  $ 10^{16}$  & 30 & 3.0 & 15 &
$0.56^{\ddag}$ & 17.3 & 22.5 \\
\hline
\end{tabular}
\end{center}

\hspace*{0.1cm}{\footnotesize $^{*}$ estimated using eq. (7), Paper I.} \\
\hspace*{1cm}{\footnotesize $^{\dag}$ estimated using eq. (2.13), Falle \&
Raga, 1993.} \\ \hspace*{1cm}{\footnotesize $^{\ddag}$ estimated using eq.
(8), Paper I.} \\

\newpage

\begin{center}
{\bf {References} \\ }
\end{center}

\setlength{\baselineskip}{18pt} \setlength{\parindent}{0cm} Bachiller, R.,
Cernicharo, J., Mart\'in-Pinatdo,J., Tafalla, M., \& Lazareff, B. \\
\hspace*{1cm}1990, {\em Astr. Ap}, 231, 174.

Benz, W.1990, in $Numerical$ $Modeling$ $of$ $Stellar$ $Pulsations$: $%
Problems$ \\ \hspace*{1cm}$and$ $Prospects$, ed. J.R. Buchler
(Dordrecht:Kluwer), p. 269.

Benz, W.1991, in $Late$ $Stages$ $of$ $Stellar$ $Evolution$ $and$ $%
Computational$ $Methods$ $in$ \\ \hspace*{1cm}$Astrophysical$ $Hydrodynamics$%
, ed. C. de Loore (Berlin: Springer), p. 259.

Birkinshaw, M. 1991, in {\em Beams and Jets in Astrophysics}, ed. P.A.
Hughes Cambridge, \\ \hspace*{1cm}University Press, p. 278.

Blondin, J.M., Fryxell, B.A., \& K\"onigl, A. 1990, {\em Ap. J.}, 360, 370
(BFK).

Bohm, K.-H. \& Solf, J. 1985, {\em Ap. J.}, 294, 533

Chernin, L., Masson, C., Gouveia Dal Pino, E.M. \& Benz, W. 1994, $Ap.$ $J.$
(in press) \\ \hspace*{1cm} (Paper II)

Dopita, M.A.1978, $Ap.$ $J.$ $Suppl$., 37, 117.

Falle, S.A.E.G. \& Raga, A.C. 1993, $MNRAS$, 261, 573.

Gouveia Dal Pino, E.M. \& Benz, W. 1993a, $Ap.$ $J.$, 410, 686 (Paper I)

Gouveia Dal Pino, E.M. \& Benz, W. 1993b, in {\em Proceedings of} $the$ $%
Sub-Arcsecond$ $Radio$ \\ \hspace*{1cm}$Astronomy$ $Manchester$ $Conference$
, eds. Davis \& Booth (Cambridge \\ \hspace*{1cm}University Press), p.

Gouveia Dal Pino, E.M. \& Benz, W. 1993c, in {\em Proceedings of the IV
International Toki \\ \hspace*{1cm}Conference on Plasma Physics and
Controlled Fusion, }eds. T.D. \\ \hspace*{1cm}Guyenne \& J.J. Hunt. (ESA\
Publs.), p. 333.

Gouveia Dal Pino, E.M., Birkinshaw, M. \& Benz, W. 1994 (in preparation).

Hartigan, P., Raymond, J., \& Meaburn, J. 1990, {\em Ap. J.}, 362, 624.

Hartigan, P., Raymond, J., \& Hartmann, L. 1987, {\em Ap. J.}, 316, 323.

Hartigan, P., \& Raymond, J. 1992, $Ap.J.$, 409, 705 (HR).

Hartmann, L., Kenyon, S., \& Hartigan, P. 1993, in {\em Protostars and
Planets III}, (University \\ \hspace*{1cm}of Arizona Press), in press.

Innes, D.E., Giddings, J.R., \& Falle, S.A.E.G. 1987, $MNRAS$, 226, 67.

Katz, J. 1989, Ph.D. thesis, Princeton University.

Kofman, L., \& Raga, A.C., 1992, {\em Ap. J.}, 390, 359 (KR).

Lizano, S. et al. 1988, {\em Ap. J.}, 328, 763.

Masson, C.R. \& Chernin, L.1993, {\em Ap. J.}, 414, 230.

Morse, J.A., Hartigan, P., Cecil, G., Raymond, J.C., \& Heathcote, S. 1992,
{\em Ap. J.},339, 231.

Morse, J.A., Heathcote, S., Cecil, G., Hartigan, P. \& Raymond, J.C. 1993, $%
A.$ $J$., in press.

Mundt, R. 1988, in {\em Proceedings of NATO-ASI on Formation and Evolution
of Low Mass} \\ \hspace*{1cm} {\em Stars}, eds. A. Dupree and M.T.V.T. Laga,
(Reidel:Dordrecht), p. 257.

Mundt, R., Br\"ugel,E.W., \& Buhrke, T. 1987, {\em Ap. J.}, 319, 275 (MBB).

Mundt, R., Buhrke, T., Fried, J.W., Neckel, T., Sarcander, M. \& Stocke, J.
1984, {\em Astr. Ap}, \\ \hspace*{1cm}140, 17

Norman, C.A. \& Silk, J. 1979, $Ap.$ $J.$, 228, 197.

Raga, A.C. 1992, {\em MNRAS}, 258, 301.

Raga, A.C. \& Cabrit, S. 1993, {\em Astr. Ap}, in press.

Raga, A.C., Cant\'o, J., Binette, L., \& Calvet, N. 1990, {\em Ap. J.}, 364,
601.

Raga, A.C., Canto, J., Calvet, N., Rodriguez, L.F. \& Torrelles, J. 1993,
{\em Astr. Ap}, 276, 539.

Raga, A.C., \& Kofman, L., 1992, {\em Ap. J.}, 386, 222 (RK).

Raga, A.C., \& Noriega-Crespo, A.1993, preprint.

Rees, M.J. 1978, MNRAS, 184, 61P.

Reipurth, B., 1985, {\em Astr. Ap}, 143, 435.

Reipurth, B., 1989a, in {\em Proceedings of the ESO Workshop on Low Mass
Star Formation and} \\ \hspace*{1cm} {\em Pre-Main Sequence Objects}, ed. B.
Reipurth, (Garching:ESO), p. 247.

Reipurth, B., 1989b, {\em Nature}, 340, 42.

Reipurth, B. Bally, J., Graham, J. A., Lane, A.P., \& Zealey, W.J. 1986, $A$
\& $A$, 164, 51.

Reipurth, B. \& Heathcote, S. 1991, {\em Astr. Ap}, 246, 511.

Reipurth, B., Raga, A.C. \& Heathcote, S. 1992, {\em Ap. J.}, 392, 145.

Rodr\'iguez, L.F. 1989, $Rev.Mexic.Astron.Astrof.$, 18, 45.

Stahler, S. 1993 in Astrophysical Jets, ed. M. Livio, C.O.'Dea, \& D.
Burgarella (Cambridge \\ \hspace*{1cm}Univ. Press).

Stone, J.M., \& Norman, M.L. 1993a, preprint (SN).

Stone, J.M., \& Norman, M.L. 1993b, preprint.

Strom, K.M., Strom, S.E., \& Stocke, J. 1983, {\em Ap. J.}, 271, L23

Wilson, M.J. 1984, {\em MNRAS}, 209, 923. \newpage

\begin{center}
{\bf {Figure Captions} \\ }
\end{center}

\setlength{\baselineskip}{24pt}

Figure 1. The mid-plane density contours evolution of an intermittent
cooling jet with intermediate period of variability ($\tau _{on}=\tau
_{off}=3(R_j/v_j)\simeq 127$ yrs), and initial parameters $\eta =10,$ $%
n_a=1000$ cm$^{-3}$, $R_j=2\times 10^{16}$ cm, $v_j=150$ km/s, $M_j=9.5$,
and $M_a=3$ (model I of Table 1). The z and x coordinates are in units of $%
R_j$. The contour lines are separated by a factor of $1.2$ and the density
scale covers the range from $\simeq 0.01$ up to $65/n_a$. The times depicted
are (in units of $R_j/c_a=127$ yr): (A) 2.4; (B) 4.1; (C) 5.9; (D) 8.3; (E)
9.2; and (F) 11.0.

Figure 2. A closer view of the last snapshot of Figure 1 at $t\simeq
11(R_j/c_a)\simeq 1387$ yr. (A) shows the central density contour, and (B)
the corresponding velocity distribution map within $y=\pm 0.3R_j$.

Figure 3. The axial density along the symmetry axis $(x=0=y)$ of the jet for
the six snapshots shown in Figure 1. The peak corresponding to the leading
working surface at the head of the jet is labeled ''$ws$''. The internal
shocks are labeled with numbers which increase according the order of
appearance of each shock. The density and the distance $d$ along the z axis
are scaled with the values $n_a=1000$ cm$^{-3}$ and $R_j=2\times 10^{16} $
cm, respectively.

Figure 4. The density (solid lines) and pressure (dashed lines) across the
flow (of the central slice) as a function of the distance along the jet of
Figure 1 at $t\simeq 11(R_j/c_a)\simeq 1387$ yrs. The four positions
depicted correspond to the four internal shocks of Figures 1F and 3F. The
pressure and density of the leading working surface are very high and have
been clipped in the figure in order to highlight the lower level features of
the internal outflow. The vertical scales can be calibrated using the
markers in the top right corner.

Figure 5. The position of the leading bow shock (labeled $ws$) and the
internal shocks at the symmetry axis of the jet as a function of time. The
position and time are in units of $R_j$ and $R_j/c_a=127$ yr, respectively.
The shocks labeled with numbers 6 and 7 have emerged in the jet inlet after $%
t=12R_j/c_a=1524$ yr when the leading working surface had already left the
top of computation domain at $z_{max}=24R_j$.

Figure 6. The density (n) and pressure (p) evolution of the internal knots
along the axis of the jet of Figure 1. The time interval depicted is the
same of Figure 5. It is given in units of $R_j/c_a=127$ yr.

Figure 7. The mid-plane density contours evolution and the corresponding
velocity distribution of an intermittent cooling jet with a long period
variability ($\tau _{on}=22.5\tau _{dy}=143$ yr, and $\tau _{off}=17.3\tau
_{dy}=110$ yrs) and input parameters $\eta =3$, $n_a=60$ cm$^{-3}$, $%
R_j=10^{16}$ cm, $v_j=500$ km/s, $M_j=30,$ and $M_a=17.3$ (model II of Table
1). The z and x coordinates are in units of $R_j$. The density contour lines
are separated by a factor of $1.2$ and the density scale covers the range
from $\simeq 0.12$ up to $1130/n_a$. The minimum velocity in the velocity
map is $\simeq 10^{-6}c_a$ and the maximum velocity is $\simeq 20c_a$, with $%
c_a=29$ km/s. The times depicted are (in units of $R_j/c_a=110$ yr): (A)
1.6; (B) 2.5; and (C) 3.4.

\end{document}